\documentclass[prl,twocolumn,showpacs,floatfix]{revtex4-1}
\usepackage{graphicx}
\usepackage{amssymb}
\usepackage{amsmath}
\usepackage{xspace}
\usepackage{url}
\usepackage{subfigure}

\newcommand{\AlTwoOThree}{Al$_{\mathrm{2}}$O$_{\mathrm{3}}$}
\newcommand{\MgO}{MgO}

\begin{document} %%%%%%%%%%%%%%%%%%%%%%%%%%%%%%%%%%%%%%%%%%%%%%%%%%%%%%%%%%
%------------------------------------------------------------------------------ 
% Title
%------------------------------------------------------------------------------
%\title{Absorption of an electron by a dielectric surface}
\title{Absorption of an electron by a dielectric wall}

%------------------------------------------------------------------------------ 
% Authors
%------------------------------------------------------------------------------
%------------------------------------------------------------------------------ 
% Date
%------------------------------------------------------------------------------
\author{F. X. Bronold and H. Fehske} 
\affiliation{Institut f{\"ur} Physik,
             Ernst-Moritz-Arndt-Universit{\"a}t Greifswald,
             17489 Greifswald,
             Germany}

\date{\today}
\begin{abstract}
We introduce a method for calculating the probability with which a low-energy 
electron hitting the wall of a bounded plasma gets stuck in it and apply the method 
to a dielectric wall with positive electron affinity smaller than the bandgap 
using \MgO\ as an example. In accordance with electron beam scattering data we 
obtain energy-dependent sticking probabilities significantly less than unity 
and question thereby for electrons the perfect absorber assumption used 
in plasma modeling. 
\end{abstract}
\maketitle

The interaction of electrons with surfaces plays a key role in applied science. 
Various methods of surface analysis~\cite{Cazaux12,SW13,Werner01} are based on 
it as well as a number of materials processing techniques~\cite{KBK14}. In 
these applications the electron energy is above 100\,eV and backscattering and 
secondary electron emission, the physical processes involved, are sufficiently 
well understood~\cite{GT03,SF02,DJT00,Vicanek99,DRW95,Tofterup85,KO72,Dashen64}. 
The situation is different for electron-surface interaction at energies below
100\,eV, as it occurs in dielectric barrier discharges~\cite{BWS12,TBW14,PS15},
dusty plasmas~\cite{Tolias14a,RBP10,Ishihara07,FIK05,Mendis02}, 
Hall thrusters~\cite{DRF03,BMP03}, and electric probe measurements~\cite{LMB07}.
Much less is quantitatively known about it, especially at very low energies, 
below 10\,eV. For instance, the probability with which a low-energy electron gets 
stuck in the wall after hitting it from the plasma is unknown. In the modeling 
of bounded plasmas~\cite{SEK13,SAB07,KDK04,BMG05,USB00} it is assumed to be close
to unity, implying for electrons the wall is a perfect absorber~\cite{Alpert65}, 
irrespective of the material. Since electron absorption and extraction (by 
charge-transferring heavy particle collisions) control the wall potential, and hence 
the plasma sheath, which in turn affects the bulk plasma, the electron sticking 
probability is a crucial parameter. That its magnitude matters and should 
be known precisely has been recognized most clearly by 
Mendis~\cite{Mendis02} in connection with grain charging in dusty plasmas but 
the theoretical work~\cite{UN80,HS70} he refers to is based on classical 
mechanics not applicable to electrons.

In this work we apply quantum mechanics to calculate the probability 
with which a low-energy electron is absorbed by a surface. We couch the presentation 
in a particular application: The calculation of the electron sticking probability at 
room temperature for a dielectric wall~\cite{Bechstedt03} with positive electron 
affinity $\chi$ smaller than the bandgap $E_g$. But the approach is general and can 
be also applied to other cases. It utilizes two facts noticed by Cazaux~\cite{Cazaux12}: 
(i) low-energy electrons do not see the strongly varying short-range potentials 
of the surface's ion cores but a slowly varying surface potential and (ii) they 
penetrate deeply compared to the lattice constant into the surface. The scattering 
pushing the electron back to the plasma occurs thus in the bulk of the wall 
suggesting the probability for the electron to get absorbed by it to be the probability 
for transmission through the wall's surface potential times the probability to stay 
inside the wall despite of internal backscattering.

\begin{figure}[t]
\includegraphics[width=\linewidth]{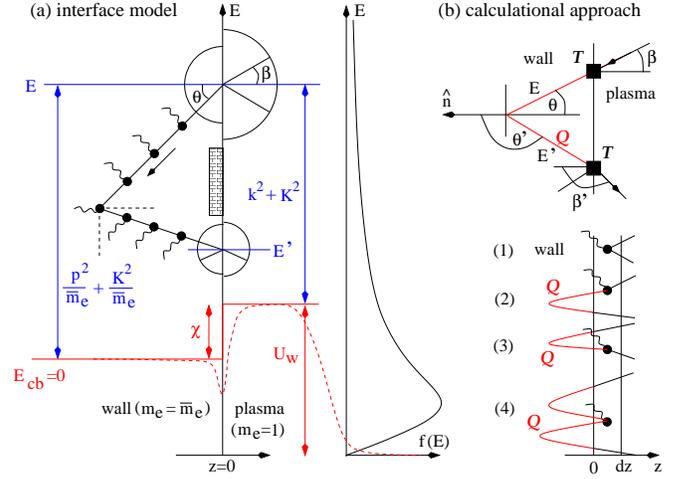}
\caption{(color online)
(a) Potential energy step of height $\chi$ modeling the wall and a scattering
trajectory bringing an electron entering the wall back to the plasma.
Half circles are the moduli of the electron momenta outside and inside
the wall at the entrance ($E$) and exit ($E^\prime$) point connected by
phonon emission events. Also shown is the Coulomb barrier
$U_{\rm w}$ the electron has to overcome to reach the wall, the energy 
distribution $f(E)$ it may have, and a realistic surface potential (red dashed curve).
The tilt of the potential step due to the wall charge responsible for $U_{\rm w}$ 
is not shown. (b) Illustration of Eqs.~\eqref{StickCoeff}--\eqref{EscapeProb}
defining $S(E,\xi)$. The
potential step leads to a transmission probability ${\cal T}$ while the emission
of phonons to a conditional backscattering probability ${\cal E}$
proportional to a quantity $Q$, obtained from the invariant embedding approach,
that is, by adding an infinitesimally thin layer to the wall
and requiring $Q$ to be invariant against it~\cite{Dashen64}. Notice the
definition of the angles and energies characterizing the electron.
Inside (outside) the wall the angles, measured with respect to the surface normal
$\hat{n}$, are denoted $\theta$ ($\beta$) and $\theta^\prime$
($\beta^\prime$). 
}
\label{Model}
\end{figure}

The variation of the potential across a floating dielectric wall can be calculated
self-consistently~\cite{HBF12}. For $\chi>0$ it gives
rise to an energy barrier for electrons whose height on the plasma side is 
the Coulomb energy $U_{\rm w}$ an electron has to overcome to reach the wall.
On the solid side the height is $\chi$ since for $\chi > 0$ electrons 
entering the wall and making up its charge are surplus electrons occupying the
wall's conduction band~\cite{HBF12}. For the sticking probability it is the
kinetic energy of the approaching electron in the vicinity of the wall which matters.
It suffices therefore to model the wall by a three-dimensional potential step 
with height $\chi$ and electron mass mismatch $\overline{m}_e=m_e^*/m_e < 1$, where
$m_e^*$ is the effective electron mass in the conduction band of the wall and $m_e$ 
is the electron mass, as illustrated in Fig.~\ref{Model}a. More refined treatments
are possible. The shape of the step however is largely irrelevant for the energies 
we consider. For \AlTwoOThree, for instance, the attenuation length for an electron
with a few eV is around $200\,\AA$~\cite{Hickmott65} and thus 
much larger than the lattice spacing setting the scale on which the step varies.

Assuming as in Fig.~\ref{Model}a the wall (plasma) to occupy the $z<0$ ($z>0$) half 
space, measuring energy in Rydbergs from the bottom of the conduction band, that is, setting 
$E_{\rm cb}=U_{\rm w}-\chi\equiv 0$, length in Bohr radii, and mass in electron masses, the 
transmission probability for a plasma electron hitting the wall, and 
having thus kinetic energy $E-\chi > 0$, is given by~\cite{WY79}
\begin{align}
{\cal T}(E,\xi)=\frac{4\overline{m}_e k p}{(\overline{m}_e k + p)^2}
\label{Trans}
\end{align}
with $k=\sqrt{E-\chi}\,\xi$ and $p=\sqrt{\overline{m}_e E}\,\eta$ the $z-$components of the
electron momenta outside and inside the wall. In~\eqref{Trans} the signs of $k$ and $p$ are 
always the same. We can thus define the direction cosines $\xi$ and $\eta$ referenced, 
respectively, to the electron momenta outside and inside the wall, by their absolute values:
$\xi=|\cos\beta|$ and $\eta=|\cos\theta|$ (see Fig.~\ref{Model}b
for the definition of $\beta$, $\theta$, and $E$).
This choice is dictated by the modeling of 
internal backscattering presented later. Since the potential varies only perpendicularly to 
the interface the lateral momentum $\vec{K}$ is conserved. Together with energy conservation,
$E=\chi+k^2+\vec{K}^2=(p^2+\vec{K}^2)/\overline{m}_e$, this yields
\begin{align}
1-\eta^2 = \frac{E-\chi}{\overline{m}_e E}\big(1-\xi^2\big)~
\label{etaxi}
\end{align}
connecting 
%the direction cosines 
$\eta$ and $\xi$. From \eqref{etaxi} follows that an 
electron from the plasma approaching the wall with kinetic energy $E-\chi=\vec{K}^2 + k^2>0$ 
enters it only when 
%$\xi$ is larger than
\begin{align}
\xi>\xi_c=\left\{\begin{array}{ll}
0 & ~{\rm for}~~ \chi < E < E_0 \\
\sqrt{1-\frac{\overline{m}_e E}{E-\chi}} & ~{\rm for}~~ E>E_0
\end{array}\right. 
\label{xic}
\end{align}
with $E_0=\chi/(1-\overline{m}_e)$. For $\xi < \xi_c$ the electron is in an evanescent 
wave with $p^2<0$, once it is inside the wall, and is thus totally reflected~\cite{GB89}. This 
fact, unnoticed so far in connection with dielectric walls, has strong implications for the 
sticking probability. In addition, Eq.~\eqref{etaxi} may force an electron entering the 
wall from the plasma to instantaneously acquire inside the wall a perpendicular kinetic energy 
$p^2/\overline{m}_e < \chi$. For $\overline{m}_e < 1$, applicable to \MgO, the material we 
use for illustration because of the availability of data from electron beam scattering 
experiments~\cite{CF62}, this happens when $\xi<\sqrt{1-\overline{m}_e}$. If such an electron 
cannot gain energy by inelastic scattering, as it is the case for dielectric walls at room 
temperature, it will have no chance to come back to the plasma since it will not be able
to overcome the potential step $\chi$. 

The transmission probability captures only the ballistic aspects of
electron absorption by the wall. Once the electron is inside the wall it is 
subject to elastic and inelastic internal scattering, which may push the electron 
back to the interface and, after successfully traversing the surface potential 
in the reverse direction, back to the plasma (see Fig.~\ref{Model}a for an illustration). 
To take this possibility into account we define the sticking probability $S(E,\xi)$ as the 
probability of an electron hitting the wall from the plasma with energy $E$ and direction 
cosine $\xi$ not to return to it after entering the wall and suffering backscattering inside it. 
Hence, 
\begin{align}
S(E,\xi) = {\cal T}(E,\xi)[1-{\cal E}(E,\xi)]~
\label{StickCoeff}
\end{align}
with the conditional probability 
\begin{align}
{\cal E}(E,\xi)=\frac{\int_{\eta_{\rm min}}^1 \!\!\!\!\!\! d\eta^\prime
\int_{E^\prime_{\rm min}}^E \!\!\!\!\!\! dE^\prime \rho(E^\prime)Q(E\eta(\xi)|E^\prime\eta^\prime)
{\cal T}(E^\prime,\xi(\eta^\prime))}
{\int_0^1 d\eta^\prime \int_0^E dE^\prime \rho(E^\prime)Q(E\eta(\xi)|E^\prime\eta^\prime)}
\label{EscapeProb}
\end{align}
for the electron to escape from the wall after at least one backscattering event. The 
lower integration limits in the numerator, $\eta_{\rm min}=\sqrt{\chi/E}$ and 
$E^\prime_{\rm min}=\chi/\eta^{\prime\, 2}$, ensure that only events are counted for 
which the perpendicular post-collision energy $p^{\prime\, 2}/\overline{m}_e>\chi$, 
$\rho(E)=\sqrt{\overline{m}_e^3E}/2(2\pi)^3$ is the conduction band density of states, 
and $Q(E\eta|E^\prime\eta^\prime)$ is proportional to the probability for an electron, 
penetrating the wall in a state with energy $E$ and direction cosine $\eta$, to backscatter 
after an arbitrary number of internal scattering events towards the interface in a state 
with energy $E^\prime$ and direction cosine $\eta^\prime$ (see Fig.~\ref{Model}b). Since 
$Q(E\eta|E^\prime\eta^\prime)$ describes backscattering, $0 < \theta \le \pi/2$ and 
$\pi/2 < \theta^\prime \le \pi$, implying 
$0\le \eta,\eta^\prime < 1$ as $\eta^\prime=|\cos\theta^\prime|$. The energy integrals 
in~\eqref{EscapeProb} anticipate that for a dielectric wall at room temperature the electron
cannot gain energy and $\eta(\xi)$ and its inverse $\xi(\eta)$ are defined by Eq.~\eqref{etaxi}.

To calculate $Q(E\eta|E^\prime\eta^\prime)$ we use the invariant embedding approach, 
originally applied to the electron backscattering problem by Dashen~\cite{Dashen64} 
and revitalized recently by Vicanek~\cite{Vicanek99} and Glazov and coworkers~\cite{GT03,GP07}.
It is illustrated in Fig.~\ref{Model}b. Summing up the four paths (1)--(4), requiring 
$Q$ to be invariant against the change it induces, and taking into account that $Q$ does 
not depend on the azimuths of the electron's initial and final propagation direction   
measured with respect to the surface normal $\hat{n}$ yields a nonlinear integral equation 
for $Q(E\eta|E^\prime\eta^\prime)$~\cite{Dashen64,GP07}. 
Since at low energy most scattering processes in a solid are forwardly peaked~\cite{Ridley99}
we linearize the equation by an expansion of $Q(E\eta|E^\prime\eta^\prime)$ 
in the number of the rare backscattering events. For dielectric surfaces at room 
temperature and $0<\chi<E_g$ scattering at low energy arises mainly from the emission of 
optical phonons with energy $\omega$. Since the electron cannot gain energy by scattering
it can loose at most the energy it initially had when entering the wall. Expanding 
$Q(E\eta|E^\prime\eta^\prime)$ also in the number of forward scattering events yields 
thus a double series which terminates after a finite number of terms. From the golden
rule transition rate per time~\cite{Ridley98} and the material parameters of
\MgO~\cite{KYA08,OGP03} follows that backward scattering due to emission of an optical 
phonon is two orders of magnitude less likely than forward scattering. Writing
\begin{align}
Q(E\eta|E^\prime\eta^\prime)=\sum_{n=1}^\infty\sum_{m=0}^\infty Q^n_{m}(E;\eta|\eta^\prime)
\delta(E-E^\prime-\omega^n_m)
\label{Qexpansion}
\end{align}
with $\omega^n_m=(n+m)\omega$ we can thus truncate the summation already after a single backward
scattering event, that is, after $n=1$ leaving $M_{\rm tot}=\lfloor E/\omega \rfloor -1$ forward 
scattering events at most. Introducing kernels
\begin{align}
K^\pm(E\eta|E^\prime\eta^\prime) &= \frac{1}{2\rho(E)}
\big[ (E+E^\prime \mp 2\sqrt{E E^\prime} \eta\eta^\prime)^2 \nonumber\\
&- 4 E E^\prime (1-\eta^2)(1-\eta^{\prime 2})\big]^{-1/2}~,
\label{Kkernel}
\end{align}
for forward ($+$) and backward ($-$) scattering and a function $\Pi(E)={\rm arcosh}(\sqrt{E/\omega})/E$,
all three arise in the course of the calculation from the transition rate per length, the input
driving the integral equation for $Q$~\cite{GP07} most directly obtained by 
dividing the transition rate per time by the electron velocity prior the emission event, 
%all three can be obtained from the transition rate per time by turning it via division by the 
%pre-emission electron velocity into a transition rate per length, 
the expansion coefficients satisfy
\begin{align}
Q^1_{m}(E;\eta|\eta^\prime) &=
F_m(E;\eta|\eta^\prime)Q^1_{m-1}(E-\omega;\eta|\eta^\prime)\nonumber\\
&+Q^1_{m-1}(E;\eta|\eta^\prime)G_m(E-m\omega;\eta|\eta^\prime)~
\label{RR}
\end{align}
with $m=1,...,M_{\rm tot}$, and
\begin{align}
F_m(E;\eta|\eta^\prime) &= \frac{K^+(E|E-\omega;\eta)\eta^\prime\rho(E-\omega)}
                        {\eta^\prime \Pi(E) + \eta\Pi(E-(m+1)\omega)}~, \\
G_m(E;\eta|\eta^\prime) &= \frac{\eta\rho(E)K^+(E|E-\omega;\eta^\prime)}
                        {\eta^\prime \Pi(E+m\omega) + \eta\Pi(E-\omega)}~.
\end{align}
The initialization of the recursion~\eqref{RR} is given by 
%the single backward scattering event,
\begin{align}
Q^1_{0}(E;\eta|\eta^\prime) = \frac{\eta^\prime K^-(E\eta|E-\omega\eta^\prime)}
{\eta^\prime \Pi(E) + \eta \Pi(E-\omega)}~.
\end{align}

Forward scattering is encoded into the function
%\begin{align}
%K^+(E|E^\prime;\eta)=\int_0^1 \!\!\! d\bar{\eta} K^+(E\eta|E^\prime\bar{\eta})
%                    =\int_0^1 \!\!\! d\bar{\eta} K^+(E\bar{\eta}|E^\prime\eta)~,$
$K^+(E|E^\prime;\eta)=\int_0^1 d\bar{\eta} K^+(E\eta|E^\prime\bar{\eta})
                    =\int_0^1 d\bar{\eta} K^+(E\bar{\eta}|E^\prime\eta),$
%\end{align}
where the second identity utilizes the symmetry of $K^+(E\eta|E^\prime\eta^\prime)$ with 
respect to interchanging $\eta$ and $\eta^\prime$. It can be obtained analytically 
and enters the formalism because $K^+(E\eta|E^\prime\eta^\prime)$ is strongly peaked for 
$\eta=\eta^\prime$, as can be seen from Eq.~\eqref{Kkernel}, and noticing that 
$K^+(E\eta|E^\prime\eta^\prime)$ enters always with $E^\prime=E-\omega$. 
Integrals over the direction cosine containing $K^+(E\eta|E^\prime\eta^\prime)$
can thus be handled by a saddle-point approximation. Physically this means  
the directional change due to forward scattering is negligible. Only backward scattering changes 
the direction (see Fig.~\ref{Model}a). We thus end up with a model similar in spirit to the Oswald, 
Kasper and Gaukler model~\cite{OKG93} for backscattering of electrons from surfaces due to 
multiple elastic scattering. The strong forward scattering comes from the matrix element for the 
interaction of an electron with an optical phonon which is at low electron density inversely 
proportional to the momentum transfer squared~\cite{Ridley98}. 
\begin{figure}[t]
\begin{minipage}{0.49\linewidth}
%\rotatebox{270}{\includegraphics[width=0.8\linewidth]{SmMgO.eps}}

\vspace{2mm}
\rotatebox{270}{\includegraphics[width=0.85\linewidth]{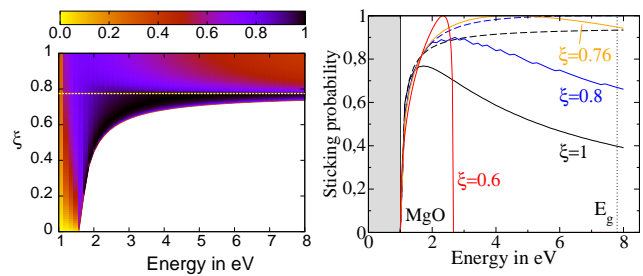}}
\end{minipage}\begin{minipage}{0.49\linewidth}

\hfill\includegraphics[width=\linewidth]{Fig2b.eps}
\end{minipage}
\caption{(color online) Electron sticking probability for 
\MgO\ obtained from Eq.~\eqref{StickCoeff}. On the left is shown
$S(E,\xi)$ for all direction cosines $\xi$ and energies $E$ up
to the bandgap $E_g$. Total reflection occurs in the white region. 
Below the yellow dotted line inelastic
backscattering has no effect on $S(E,\xi)$. On the right
$S(E,\xi)$ (solid curves) and ${\cal T}(E,\xi)$ (dashed curves) are plotted 
as a function of $E$ for representative $\xi$. The material parameters are
$\overline{m}_e=0.4$, $\chi=1\,{\rm eV}$, 
$E_g=7.8\,{\rm eV}$~\cite{KYA08}, and $\omega=0.1\,{\rm eV}$~\cite{OGP03}. 
The value for $\overline{m}_e$ is strictly 
applicable only at the band edge. Away from it $\overline{m}_e$ depends on 
momentum. 
%We ignore this dependence. 
From the bandstructure~\cite{XC91}  
we expect our results reliable up to $5$\,eV enough 
for a comparison with experiment (see Fig.~\ref{ExpData}).}
\label{AngleResolved}
\end{figure}

Substituting expansion~\eqref{Qexpansion} for $Q(E\eta|E^\prime\eta^\prime)$ in~\eqref{EscapeProb}
sets $E^\prime$ to $E_m^1=E-\omega_m^1$ and replaces the energy integrals by sums over 
$m$ running in the numerator up to $m=M_{\rm open}=\lfloor (E\eta^2-\chi)/(\eta^2\omega) \rfloor - 1$ 
and in the denumerator up to $m=M_{\rm tot}$. Inserted into Eq.~\eqref{StickCoeff} it finally gives 
the electron sticking probability $S(E,\xi)$ at room temperature for a dielectric wall 
with $0 < \chi < E_g$ and a clean, lateral momentum conserving interface.

In Fig.~\ref{AngleResolved} we present the results for \MgO\ obtained from  Eq.~\eqref{StickCoeff}.
Material parameters are given in the caption. The left panel shows the angle-resolved sticking
probability $S(E,\xi)$ over the whole range of direction cosines and energies
up to the bandgap, above which electron-hole pair generation across the
bandgap has to be taken into account as an additional, Coulomb-driven backscattering 
process. It can be treated in the same spirit, in particular the truncation after 
one backscattering event will still be possible, but the recursion will then
contain energy integrals. The white area in the plot for $S(E,\xi)$ indicates the 
region in the $(E,\xi)$-plane where total reflection occurs. It is large because 
$\overline{m}_e$ is for \MgO\ significantly smaller than unity. Below the dotted yellow 
line, $\xi=\sqrt{1-\overline{m}_e}$, inelastic backscattering due to emission 
of optical phonons is irrelevant for sticking because the perpendicular energy of 
the electron drops, upon entering the wall, below the potential step $\chi$ due to 
conservation of total energy and lateral momentum. It is hence already confined by 
transmission. Only above the dashed yellow line inelastic backscattering has a chance to 
bring the electron, once it is inside the wall, back to the interface and, after a successful 
reverse transmission through the surface potential, back to the plasma. As a result, 
$S(E,\xi)={\cal T}(E,\xi)$ for $\xi < \sqrt{1-\overline{m}_e}$ and $S(E,\xi)<{\cal T}(E,\xi)$ 
for $\xi > \sqrt{1-\overline{m}_e}$. This can be more clearly seen in the right panel,
where $S(E,\xi)$ (solid curves) and ${\cal T}(E,\xi)$ (dashed curves) are 
plotted as a function of $E$ for representative $\xi$. 

%Compared to Monte Carlo simulations~\cite{SW13,Werner01,DJT00} the recursion is 
%very efficient. For instance, the calculation of the about 1500 data points for 
%$S(E,\xi)$ shown in the left panel of Fig.~\ref{AngleResolved}, corresponding each 
%to a sum of trajectories with one backward and (depending on energy) up to 78 
%forward scattering events, with the former interlaced between the latter in all 
%possible ways, requires only one hour computing time on a notebook.

The sticking probabilities in Fig.~\ref{AngleResolved} are for $\xi<\xi_c$ 
strongly affected by total reflection caused by the mass mismatch and  
conservation of energy and lateral momentum. For the latter 
in-plane homogeneity of the interface is crucial. In reality imperfections 
may destroy it and lateral momentum may thus not be conserved. To take this possibility
into account we include hard core potentials to mimic interfacial scattering centers 
and adopt the modeling of ballistic electron emission spectroscopy by 
Smith and coworkers~\cite{SLN98} to the calculation of $S$. 
For a disordered interface we then obtain 
\begin{align}
\overline{S}(E,\xi) &= \frac{{\cal T}(E,\xi)}{1+C/\xi}[1-\overline{\cal E}(E,\xi)] \nonumber\\
                    &+ \frac{C/\xi}{1+C/\xi} \int^1_{\xi_c} d\xi^\prime
                    T(E,\xi^\prime)[1-\overline{\cal E}(E,\xi^\prime)]~,
\label{Dirty1}
\end{align}
where $\overline{\cal E}(E,\xi)$ is given by \eqref{EscapeProb} with 
${\cal T}(E,\xi)$ replaced by 
\begin{align}
\overline{\cal T}(E,\xi)=\frac{{\cal T}(E,\xi)}{1+C/\xi}  
                        + \frac{C/\xi}{1+C/\xi} \int^1_{\xi_c} d\xi^\prime {\cal T}(E,\xi^\prime)~.
\label{Dirty2}
\end{align}
The parameter $C$ is a fit parameter proportional to the density of scattering centers.
It controls the elastic scattering at the interface. In the weak scattering limit, $C\rightarrow 0$
and $\overline{S}\rightarrow S$. We thus recover the sticking probability for the clean interface.
In the strong scattering limit, $C\rightarrow\infty$ and we get the sticking probability 
for the totally disordered interface for which $C$ is irrelevant. 
\begin{figure}[t]
\includegraphics[width=\linewidth]{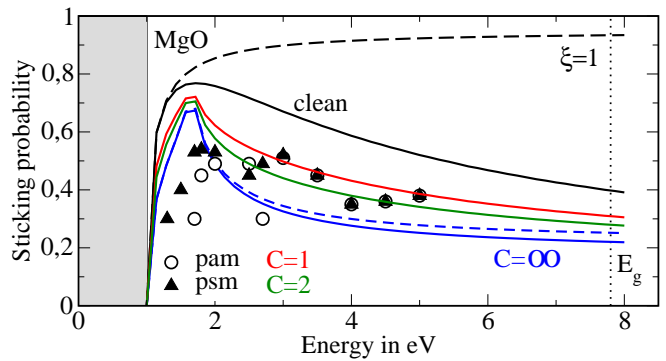}
\caption{(color online) Electron sticking probability $\overline{S}(E;1)$ for normal 
incident onto a \MgO\ wall obtained from Eq.~\eqref{Dirty1} (solid curves). We 
show data for $C=0$ (black), $1$ (red), $2$ (green), and $\infty$ (blue) with
$\overline{\cal T}(E;1)$ also included for $C=0$ and $C=\infty$ (dashed curves).
Symbols are data from electron beam scattering experiments, where two methods,
denoted potential adjusted method (pam) and potential subtraction method (psm),
have been employed to determine the energy of the electron~\cite{CF62}.
Material parameters as in Fig.~\ref{AngleResolved}.
}
\label{ExpData}
\end{figure}

Results for \MgO\ obtained from Eq.~\eqref{Dirty1}  
are shown in Fig.~\ref{ExpData} for $\xi=1$ (normal incident) and 
$C=0, 1, 2,$ and $\infty$. We also plot experimental data 
from electron beam scattering~\cite{CF62}. The agreement between
calculated and measured probabilities is astonishing indicating our approach 
captures essential aspects of electron absorption by the wall at low 
energies. Dashed curves show $\overline{\cal T}(E,1)$, the 
sticking probability in the absence of internal backscattering. For $C=0$, 
$\overline{S}(E,1)$ deviates strongly from $\overline{\cal T}(E,1)$ (black curves), 
whereas for $C=\infty$ the two quantities approach each other (blue curves). 
The reason is the angle-averaging at the disordered interface 
which lessens, for fixed $\xi$, the impact of internal backscattering compared
to the knock-out of $\xi$-values by total reflection signalled by 
the kink at $E=E_0$. How close the two curves come to each other,
that is, how ineffective internal backscattering becomes
depends on $\overline{m}_e$ and $\chi$.
%The mass mismatch $\overline{m}_e$ turns thus out to control $\overline{S}$. 
%This is not surprising. Because 
%it is the effective electron mass which subsumes at low energy the elastic scattering of 
%the electron by the ion cores of the wall.

To summarize we presented a quantum-mechanical method to calculate the sticking 
probability $S$ for a low-energy electron hitting the wall of a bounded plasma. 
%In contrast to previous work
%it is based on the quantum mechanics of electron-wall interaction. 
Using the electron's large penetration depth the method expresses $S$ as the
probability ${\cal T}$ for transmission through the wall's surface potential times the
probability $1-{\cal E}$ to stay inside it despite of internal backscattering.
For \MgO\ we got in agreement with electron beam scattering data $S<1$,
indicating $S\approx 1$, employed in plasma modeling, may not always 
be justified. Using $S(E,\xi)$ from our approach instead of $S\approx 1$, for instance, 
in the theory of grain charging in dusty plasmas would reduce the charge of micron-sized 
\MgO\ particles by more than ten percent. The method is flexible and numerically efficient. 
Various scattering processes, including interfacial scattering due to disordered 
interfaces, can be treated. It could thus guide an experimental effort to develop a realistic
description of electron-wall interaction as it occurs in many applications of bounded
plasmas.\\

We acknowledge support by the Deutsche Forschungsgemeinschaft through 
the Transregional Collaborative Research Center SFB/TRR24.

%\bibliography{./ref} 
%\bibliographystyle{apsrev}

\end{document}